\documentstyle[aps,prb,psfig]{revtex}
\begin{document}
\flushbottom
\twocolumn[
\hsize\textwidth\columnwidth\hsize\csname @twocolumnfalse\endcsname

\title{Universal conductance fluctuations and low temperature 1/f noise in
mesoscopic AuFe spin glasses}
\author{G. Neuttiens, C. Strunk\footnote{Present address: Institut f\"{u}r
Physik, Universit\"{a}t Basel, Klingelbergstrasse 82, CH-4056 Basel,
Switzerland}, C. Van Haesendonck, and Y. Bruynseraede}
\address{Laboratorium voor Vaste-Stoffysica en Magnetisme, Katholieke
Universiteit Leuven, B-3001 Leuven, Belgium}
\date{May 19, 1999}
\maketitle
\begin{abstract}
We report on intrinsic time-dependent conductance fluctuations observed in
mesoscopic AuFe spin glass wires. These dynamical fluctuations have a
$1/f$-like spectrum and appear below the measured spin glass freezing
temperature of our samples. The dependence of the fluctuation amplitude on
temperature, magnetic field, voltage and  Fe concentration allows a
consistent interpretation in terms of quantum interference effects which
are sensitive to the slowly fluctuating spin configuration.

\end{abstract}

]
The low field magnetic susceptibility of a spin glass \cite{mydosh} shows a
sharp peak near the freezing temperature $T_{f}$. Below $T_{f}$ the
magnetic impurity spins gradually freeze into random directions. The
magnetization contains a $1/f$ noise component \cite{kogan} which appears
in the vicinity of $T_{f}$ and saturates below the freezing temperature.\cite{ocio,reim} The resistance of small spin glass samples also contains
a $1/f$ noise component related to the slow dynamics of the frozen spins.\cite{bouchiat} The resistance noise may appear because of electron
quantum interference effects which are sensitive to the slow fluctuations
of the magnetic impurity configuration in the spin glass phase.\cite{feng,AS}

Quantum interference effects give rise to universal conductance
fluctuations (UCF) which for a stable defect configuration induce
reproducible fluctuations of the magnetoconductance (magnetofingerprint).\cite{washburn} In a sample having dimensions comparable to the phase coherence length $L_{\varphi}$, the fluctuation amplitude is of the order
of the conductance quantum $e^{2}/h$.\cite{LS} In larger samples, a slow
stochastic averaging of the UCF occurs. For sufficiently small non-magnetic
samples switching of a defect between two stable configurations (two-level
system) gives rise
to a UCF induced telegraph noise signal.\cite{zimmer} For larger
non-magnetic samples superposition of telegraph noise signals results in a
$1/f$ noise spectrum.\cite{giordano,birge} In mesoscopic spin glasses the
UCF will be largely destroyed by the spin flip scattering in the
paramagnetic phase above $T_{f}$. Below $T_{f}$ the dramatic slowing down
of the spin glass dynamics should allow the experimental observation of a
UCF induced noise signal.\cite{feng}    

Israeloff {\em et al.} \cite{isr-film} have measured the $1/f$ electrical
noise in CuMn spin glass films with a Mn content between 4.5 and $19.5 \,
{\rm at}.\%$. The noise amplitude shows a rapid increase near $T_{f}$
followed by a saturation at lower temperatures which is interpreted in
terms of the UCF induced noise mechanism. In smaller, mesoscopic samples
the noise signal strongly deviates from the usual Gaussian statistics.\cite{isr-mes}
The resulting spectral wandering of the noise spectrum
favors a description of the spin glass dynamics in terms of an hierarchical
model with correlated fluctuations.  Similar experiments by Meyer and Weissman on AuFe samples reveal deviations from both the droplet model and the hierarchical model for mesoscopic sample sizes.\cite{meyer} Measurements by de Vegvar {\em et al.}
\cite{vegvar} on mesoscopic CuMn wires with a Mn concentration of $0.1 \,
{\rm at.\%}$ indicate the presence of a magnetofingerprint which is stable
in time. The fingerprint is strongly altered after heating the samples to temperatures well above $T_{f}$. According to the authors this supports the idea that the UCF are sensitive to the
specific frozen spin configuration. Very recently, Jaroszy\'{n}ski {\em et
al.} \cite{dietl} have observed a $1/f$ noise signal in heavily doped
Cd$_{1-x}$Mn$_{x}$Te spin glass wires with a Mn concentration $x = 0.02$
and $x = 0.07$. The $1/f$ noise in the dilute magnetic semiconductors is
consistent with the presence of UCF induced fluctuations. The onset of the
$1/f$ noise signal coincides with the bulk $T_{f}$ value, while typical
spin glass properties such as aging and irreversibility are clearly
present. For the Cd$_{1-x}$Mn$_{x}$Te spin glass compounds the spectral
wandering of the noise spectrum rather favors an interpretation in terms of
uncorrelated droplet excitations.  

In this paper we report on high resolution measurements of the electrical
noise in small samples of the archetypical spin glass AuFe with Fe
concentrations of 0.85 and $5 \, {\rm at}.\%$. The spin flip scattering at the Fe impurities largely destroys the static magnetofingerprints. We are able to detect an excess $1/f$ noise signal whose amplitude rapidly grows at lower temperatures. Both the temperature and current dependence of the $1/f$ noise are in agreement with UCF
reflecting the dynamics of the impurity spin configuration. Our $1/f$ noise is strongly suppressed at the elevated measuring currents which have been used by Israeloff {\em et al.}\cite{isr-film,isr-mes} The low frequency noise in the AuFe spin glasses can be observed because of a dramatic slowing down of the spin dynamics due to the freezing process.  

We have performed detailed measurements of the electrical noise in narrow
AuFe spin glass wires as well as in a pure Au test wire. Table 1 gives the
relevant parameters for the samples which have been studied. The narrow
wires are obtained by flash evaporation of small pieces of a AuFe mother
alloy in resist profiles defined by electron beam lithography, followed by
lift-off. For the pure Au sample thermal evaporation of $99.9999 \, \%$
pure Au has been used. Secondary ion mass spectroscopy (SIMS) measurements
indicate that distillation effects occurring 
\begin{figure}[t]
\centerline{\psfig{file=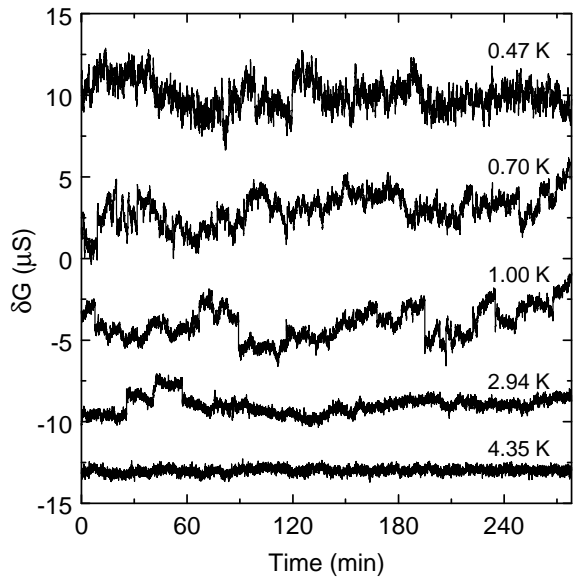,width=7.6cm}}\bigskip\bigskip
\protect\caption{ \label{Fig.1}
Time dependent fluctuations of the conductance in a $5 \,
{\rm at}.\%$ mesoscopic AuFe structure (sample W4 in Table 1). The data
are obtained by subsequent cooling of the sample towards lower
temperatures, i.e., without cycling through the spin glass freezing
temperature $T_{f}$.}
\end{figure}
\noindent during the AuFe flash
evaporation are negligible. The absence of distillation effects is
confirmed by the temperature dependence of the spin glass resistivity
\cite{neut} as well as by the temperature dependence of the anomalous Hall
resistivity \cite{vloebhall} in thicker AuFe films (see also below). 
The noise experiments have been performed with a five-terminal bridge configuration and an ac measuring current of a few kHz. A transformer (100:2000 winding ratio) cooled with liquid helium amplifies the voltage fluctuations produced by the sample and at the same time adapts the sample impedance to obtain an optimum noise figure for detecting the sample voltage with a lock-in amplifier (PAR 124A).
We are able to reliably detect voltage variations having a root-mean-square
(rms) amplitude of only $0.1 \ {\rm nV}$.   

In Fig.~\ref{Fig.1} we show the time dependence of the conductance
fluctuations which appear in a $5 \, {\rm at} . \%$ AuFe sample (sample W4
in Table 1) at different temperatures. For the measurements a $1 \, {\rm
s}$ cut-off filter has been used, implying that fluctuations with a higher
frequency are filtered out. At $T = 0.47 \, {\rm K}$, the peak to peak
variations of the conductance noise correspond to $0.1 \, e^{2}/h$. This is
a first hint which supports an interpretation in terms of UCF which are
coupled to the slow dynamics of the impurity spins below $T_{f}$.\cite{feng} 
The additional step like changes of the conductance, which
become visible at $T = 1.00 \, {\rm K}$ and $T = 2.94 \, {\rm K}$ in
Fig.~\ref{Fig.1}, may be linked to the thermally induced motion of spin
clusters.\cite{mydosh} 
      
\begin{figure}
\centerline{\psfig{file=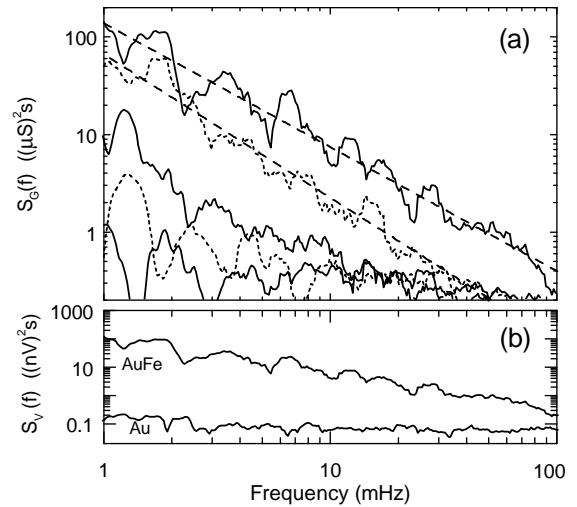,width=7.4cm}}\bigskip\bigskip
\protect\caption{(a) Noise spectra corresponding to the data shown in
Fig.~\ref{Fig.1} at temperatures $T = 0.47, 1.00, 2.94, 7.20, 12.3 \, {\rm
K}$ from top to bottom. The dashed curves correspond to a $1/f^{\alpha}$
dependence (see text). (b) Comparison of the voltage noise spectra at $T =
0.47 \, {\rm K}$ for the $5 \, {\rm at} .\%$ sample and for a pure Au
sample (sample W1 in Table 1).}
\label{Fig.2}
\end{figure}

In Fig.~\ref{Fig.2}(a) the noise power spectra $S_G(f)$ corresponding to
the data in Fig.~\ref{Fig.1} have been plotted on a double logarithmic scale. 
The low frequency noise rapidly grows 

\noindent at lower temperatures. Below $1 \, {\rm K}$,
the noise spectra can be fitted to a $1/f^{\alpha}$ dependence
indicated by the dashed lines in Fig.~\ref{Fig.2}(a). The exponent $\alpha
\simeq 1.5$ for $T = 1.00 \, {\rm K}$ and decreases towards $\alpha \simeq
1.3$ for $T = 0.47 \, {\rm K}$. At higher temperatures, the $1/f^{\alpha}$
dependence is on average still present, but an accurate determination of
$\alpha$ is not possible for the available time window. Averaging noise
spectra for different cooling cycles should be avoided in view of the
sensitivity to the particular frozen spin glass state (see also below).
Above $5 \, {\rm K}$ the noise spectra become independent of frequency and
temperature and are governed by external noise sources. 
In Fig.~\ref{Fig.2}(b) we compare the voltage noise spectra $S_V(f)$ at $T
= 0.47 \,
{\rm K}$ for the $5 \, {\rm at}. \%$ AuFe sample and a pure Au test sample
of comparable dimensions (sample W1 in Table 1). For the pure Au sample no
excess $1/f$ noise can be detected within our measuring sensitivity.
An excess noise signal is also clearly present at lower temperatures for
the AuFe samples having an Fe concentration of $0.85 \, {\rm at.} \%$.
Again, the noise  rapidly grows at lower temperatures $T < 1 \, {\rm K}$
and can be described by a $1/f^{\alpha}$ dependence with $\alpha$ in the
vicinity of 1.

\begin{figure}
\centerline{\psfig{file=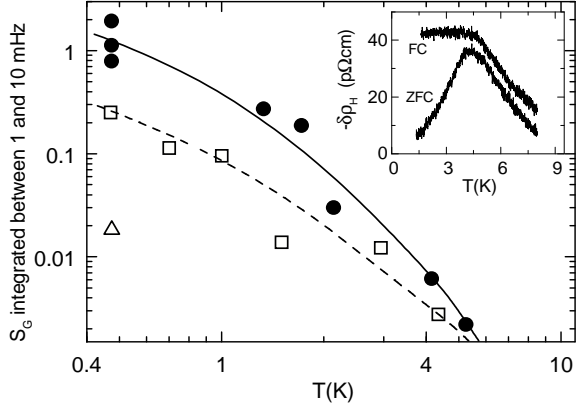,width=7.6cm}}\bigskip\bigskip
\protect\caption{Temperature dependence of the integrated conductance noise
power for the $5 \, {\rm at} .\%$ AuFe sample $(\Box)$ shown in
Fig.~\ref{Fig.1} and Fig.~\ref{Fig.2}(b) as well as for a $0.85 \, {\rm at}
.\%$ sample ($\bullet$) (sample W2 in Table 1). Both samples have the same
dimensions. The curves through the data points are only a guide to the eye.
For comparison the integrated noise power is also shown for a wider $0.85
\, {\rm at} .\%$ sample $(\triangle)$ (sample W3 in table 1). The inset
shows the temperature dependence of the Hall resistivity for a $0.85 \,
{\rm at} .\%$, $3 \, {\rm mm}$ wide film measured for field cooled (FC) and
for zero field cooled (ZFC) conditions, respectively.} \label{Fig.3}
\end{figure}

In Fig.~\ref{Fig.3} we compare the temperature dependence of  
\begin{table}
\protect\caption{Relevant parameters for the AuFe wires with different Fe
concentration $c$: Length $L$, width $w$, thickness $t$, resistivity $\rho$
and elastic mean free path $l_{el}$.} \label{Table1}
\begin{tabular}{c c c c c c c c}
Sample & $c$(at.\%) & $L$($\mu$m) & $w$(nm)  & $t$(nm)  
& $\rho$($\mu\Omega$cm) & $l_{el}$(nm) \\ \hline W1 & 0 & 1.48 & 184 & 30 &
3.15 & 26.7 \\ W2 & 0.85  & 1.46  & 187  & 23  & 13.1  & 6.44  \\
W3 & 0.85  & 7.82  & 752  & 23  & 13.5  & 6.24  \\ W4 & 5  & 1.49 &
170 & 35 & 34.3 & 2.45 \end{tabular}
\end{table}

\noindent the integrated noise power for the $5 \, {\rm at.} \%$ sample and the $0.85 \, {\rm at}.\%$ sample with comparable dimensions (sample W2 in Table 1). The plotted noise
powers have been integrated between 1 and $10 \, {\rm mHz}$ and have been
corrected for the extrinsic white background noise. The integrated noise
power has a comparable temperature dependence for both Fe concentrations,
but is larger in the sample with the lower Fe concentration. The increase of the noise
power at lower temperatures can be linked to an enhancement of the phase
coherence length $L_{\varphi}$.\cite{birge} While inelastic scattering at phonons and the other electrons becomes less effective at lower temperatures, an additional increase of $L_{\varphi}$ is caused by a reduction of the spin flip scattering rate due to the spin glass freezing process. This reduction of the spin flip scattering at lower temperatures is confirmed by a decrease of the spin glass resistivity below the freezing temperature.\cite{neut} The larger noise amplitude in Fig.~\ref{Fig.3} for the $0.85 \, {\rm at} .\%$ sample can be explained by a reduced spin flip scattering rate due to the smaller Fe content. As pointed out by Jaroszy\'{n}ski {\em et al.},\cite{dietl} the emergence of the low frequency noise requires that the spin glass dynamics, which couples to the UCF, has become sufficiently slow, with characteristic relaxation rates corresponding to our experimental measuring frequencies.

In order to be sure that the pronounced increase of the conductance noise
below $5 \, {\rm K}$ is indeed related to the spin glass freezing, we have
monitored the freezing process via measurements of the anomalous Hall
effect.\cite{vloebhall} The inset of Fig.~\ref{Fig.3} shows the
temperature dependence of the Hall resistivity for field cooled (FC) as
well as for zero field 
\begin{figure}
\centerline{\psfig{file=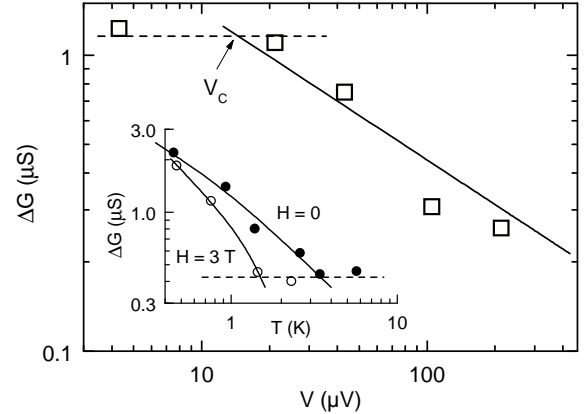,width=7.6cm}}\bigskip\bigskip
\protect\caption{Reduction of the rms conductance noise amplitude when
increasing the voltage applied across the $5 \, {\rm at} .\%$ AuFe sample
(sample W4 in Table 1). The full line corresponds to the stochastic
averaging $\propto V^{-1/2}$ which is expected to occur above the treshold
voltage $V_{c}$ (see text). The inset illustrates the reduction of the rms
conductance noise amplitude when applying a $3 \, {\rm T}$ magnetic field under field cooled conditions for a $0.85 \, {\rm at} .\%$ AuFe sample (sample W2 in Table 1). The full curves in the inset are only a guide to the eye, while the dotted line indicates the extrinsic white noise level.} 
\label{Fig.4}
\end{figure}

\noindent cooled (ZFC) measuring conditions. The data have been obtained for a 
$0.85 \, {\rm at} .\%$ AuFe film which is about $3 \,{\rm mm}$ wide and has been deposited simultaneously with the samples W2
and W3 (see Table 1). From the ZFC data we obtain a freezing temperature
$T_{f} \simeq 4.4 \, {\rm K}$ which is considerably smaller than the bulk
value $T_{f} \simeq 7.8 \, {\rm K}$. The reduction of $T_{f}$ can be linked
to finite-size scaling effects.\cite{vloebhall} 
Although $T_{f} \simeq \, 17 \, {\rm K}$ is considerably larger for the $5 \,
{\rm at} .\%$ films ($T_{f} \simeq \, 22 \, {\rm K}$ for the bulk alloy), the
temperature dependence of the integrated noise power in Fig.~\ref{Fig.3} is
similar for the $5 \, {\rm at} .\%$ sample and the $0.85 \, {\rm at} .\%$
sample, in contrast to the results obtained by Israeloff {\em et al.}
\cite{isr-film} for CuMn alloys. Unlike these authors, we also do not find
any evidence for a saturation of the $1/f$ noise signal at lower
temperatures.  

In Fig.~\ref{Fig.3} we have included the integrated noise power for a wider
and longer $0.85 \, {\rm at} .\%$ AuFe sample (sample W3 in Table 1) at the
lowest measuring temperature ($T = 0.47 \, {\rm K}$). For sample sizes exceeding the phase coherence length $L_{\varphi}$ (see below), stochastic self-averaging implies that the UCF amplitude scales with the inverse of the square root of the sample volume.\cite{washburn} Consequently, the integrated noise power should scale with the inverse of the sample volume.\cite{birge} Our experiments indicate a reduction by a factor of 8.7, while theory predicts a reduction by a factor of 6.4. 

While turning on a magnetic field of $3 \, {\rm T}$ below $T_{f}$ leaves
the noise amplitude unchanged, field cooling in the presence of a $3 \,
{\rm T}$ field delays the increase of the spin glass noise above the white
background noise. This is illustrated in the inset of Fig.~\ref{Fig.4} for
the sample W2 (see Table 1). In contrast to Fig.~\ref{Fig.3}, the white
background noise (corresponding to the dotted line) has not been subtracted
from the data points in the inset of Fig.~\ref{Fig.4}.  A shift of
the noise onset towards lower temperatures was observed before in CuMn \cite{isr-film} and in AuFe \cite{meyer} samples. For the CuMn samples,\cite{isr-film} a dependence on field history similar to ours was reported. A suppression of the noise amplitude, which depends on the magnetic field applied during thermal cycling, supports the intrinsic spin glass origin of
the excess $1/f$ noise.\cite{mydosh} In contrast to noise experiments in non-magnetic Bi samples,\cite{birge} we do not observe any reproducible magnetofingerprints. The coupling between the UCF and the fluctuating spin configuration is sufficiently strong in our samples to induce a complete scrambling of the magnetofingerprints.

An additional important piece of evidence in favor of the interpretation of
the excess noise in terms of UCF is provided by the strong reduction of the
noise signal when increasing the measuring current. This is illustrated in
Fig.~\ref{Fig.4} for the $5 \, {\rm at} .\%$ sample (sample W4 in Table 1)
at $T = 0.47 \, {\rm K}$. The data points have in this case again been
corrected to take into account the current independent white background
noise. The finite voltage across the sample induces an additional
stochastic averaging proportional to $(E_{c}/eV)^{1/2}$ (see the discussion of Fig.~12 in,\cite{washburn}) with $E_{c} = e V_{c} = \hbar D/ L_{\varphi}^{2}$ the Thouless energy and $D$ the diffusion constant. The full line in Fig.~\ref{Fig.4} corresponds to this theoretically expected reduction of the UCF at sufficiently large voltages. From the saturation at low voltages (dashed line) we infer a value for the Thouless energy $E_{c}
\simeq 0.01 \, {\rm meV}$, corresponding to a phase coherence length
$L_{\varphi} \simeq 0.3 \, \mu {\rm m}$.  Due to the spin flip scattering, $L_{\varphi}$ is about an order of magnitude smaller than for the pure Au sample. On the other hand, $L_{\varphi}$ is about five times smaller than the sample length, but remains larger than the sample width.  Taking into account the stochastic self-averaging of the UCF,\cite{washburn} the rms conductance noise amplitude for the AuFe sample ($0.03 \, e^{2}/h$, see  Fig.~\ref{Fig.4}) is about three times smaller than the rms amplitude of the magnetoconductance fluctuations in the pure mesoscopic Au sample at $T = 0.47 \, {\rm K}$ ($0.2 \, e^{2}/h$). This indicates that a considerable fraction of the UCF induced noise power is contained within the frequency range of our measurements.      

The results shown in Fig.~\ref{Fig.4} confirm that the UCF which cause the
excess noise can only be observed for very small measuring currents.
Israeloff {\em et al.} \cite{isr-film,isr-mes} have used measuring current
densities which are about two orders of magnitude larger than in our case.
This implies that their $1/f$ noise signal may have been strongly
suppressed by electron heating effects.

Finally, we note that the conductance of our samples is always much larger
than $e^2/h$, i.e., our samples reveal a pronounced metallic character.
Jaroszy\'{n}ski {\em et al.} \cite{dietl} have studied doped magnetic
semiconductors which are very close to the metal-insulator transition. This
results in a strong enhancement of the resistance noise amplitude (allowing
to observe aging and hysteresis effects), but at the same time makes it
more difficult for these authors to compare different samples. The noise
properties are, however, remarkably similar, supporting a common origin
of the $1/f$ noise for both experiments. 

In conclusion, we have observed intrinsic $1/f$ noise in narrow AuFe wires
which can be directly related to the spin glass freezing process. Our
results support the idea that the noise is caused by the time dependence of
the universal conduction fluctuations. The noise can be observed below
the freezing temperature $T_{f}$, provided the electron phase coherence length
is sufficiently large and the spin dynamics is sufficiently slow. 

We thank R. Wengerter from Vacuumschmelze GmbH for providing the core of
the cryogenic transformer. We are also much indebted to J. Vlekken of the
Limburgs Universitair Centrum for the SIMS measurements.
This work has been supported by the Fund for Scientific Research - Flanders
(FWO) as well as by the Flemish Concerted Action (GOA) and the Belgian
Inter-University Attraction Poles (IUAP) research programs. 

\end{document}